\documentclass[prl,aps,twocolumn,showpacs,amsmath,amssymb]{revtex4}
\usepackage{graphicx}
\usepackage{dcolumn}
\usepackage{bm}
\usepackage{times}

\begin{document}

\title{Spectroscopy of the Kondo Problem in a Box}

\author{Ribhu K. Kaul,$^{1,2}$ Gergely Zar\'and,$^{3,4}$ Shailesh Chandrasekharan,$^{1}$ Denis Ullmo,$^{1,5}$ and Harold U. Baranger$^{1}$}
\affiliation{
$^1$ Department of Physics, Duke University, Durham, North Carolina 27708, USA\\
$^2$ Institut f\"ur Theorie der Kondensierten Materie, Universit\"at Karlsruhe, 76128 Karlsruhe, Germany\\
$^3$ Research Institute of Physics, Technical University Budapest, Budapest, H-1521, Hungary\\
$^4$ Institut f\"ur Theoretische Fesk\"orperphysik, Universit\"at Karlsruhe, 76128 Karlsruhe, Germany\\
$^5$ Laboratoire de Physique Th\'eorique et Mod\`eles Statistiques, Universit\'e Paris-Sud, Orsay, France
}
\date{\today}

\begin{abstract}
Motivated by experiments on double quantum dots, we study the problem of a single magnetic impurity confined in a finite metallic host. We prove an exact theorem for the ground state spin, and use analytic and numerical arguments to map out the spin structure of the excitation spectrum of the many-body Kondo-correlated state, throughout the weak to strong coupling crossover. These excitations can be probed in a simple tunneling-spectroscopy transport experiment; for that situation we solve rate equations for the conductance.

\end{abstract}

\pacs{73.23.Hk, 73.21.La, 72.10.Fk}

\maketitle

An impurity spin ${\mathcal S}$ coupled anti-ferromagnetically to electrons is one of the most fundamental many-body systems, the Kondo problem \cite{Hew}. In the classic version, a ${\mathcal S}= 1/2$ impurity is coupled with anti-ferromagnetic coupling ${\mathcal J}$ to an infinite reservoir of conduction electrons with density of states $\rho$. At low temperatures, the spin is screened: its entropy is quenched by the formation of a singlet with the conduction electrons below the Kondo temperature, $T_K$, given approximately by $T_K \approx D\; e^{-1/{\mathcal J}\rho}$ for small values of $\mathcal J$, with $D$ the bandwidth of the conduction electrons.

When the size of the electronic reservoir is finite, as, {\em e.g.}, in ultra-small metallic particles \cite{DelRal,ThiSim}, electron-hole excitations cost a minimum energy of order the mean level spacing $\Delta=1/\rho$. When $\Delta \sim T_K$, the level spacing interferes with the screening of the impurity, leading to new physics \cite{ThiSim}. A finite electronic reservoir can also occur in mesoscopic semiconductor structures \cite{Marcus,David}: a small quantum dot acts as a spin $\mathcal{S}=1/2$ impurity and is coupled to a large quantum dot with a finite level spacing comparable to $T_K$. In these latter structures, the ratio $T_K/\Delta$ can be tuned by varying the voltage on the gates separating the two dots. 

In this paper we elucidate the manifestations of the strongly correlated Kondo state in the excitation spectrum of these finite size systems. The excitations can easily be obtained through measurements of the tunneling spectrum in both semiconductor systems \cite{David} and metallic nano-particles \cite{DelRal}. We provide an exact {\em theorem} on the ground state spin of the system for arbitrary coupling, and make specific statements on the excitation spectrum. Tuning $T_K/\Delta$ provides a way to study the parametric evolution of the system, a standard topic in mesoscopic physics \cite{AlhassidRMP00} but now in a many-body context, through the full weak to strong coupling crossover. In the special case of uniform level spacing \cite{ThiSim}, we find the parametric evolution of the first excitation energy, $\delta E = \Delta\; F(T_K/\Delta)$, where the universal function $F$ depends only on the parity of electrons in the reservoir.

\begin{figure}[b]
\includegraphics[width=2.1in]{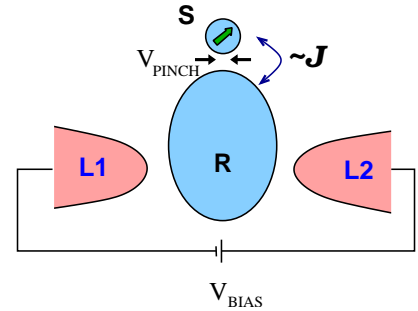}
\caption{(color online) 
A double dot system, coupled very weakly to leads (``L1'' and ``L2''). In the Coulomb blockade regime, the small dot ``S'' behaves as a spin $\mathcal{S}$ that is coupled to a finite reservoir ``R'' provided by the large dot. The leads are used to measure the excitation spectrum of the system.} 
\label{fig:setup}
\end{figure}

For concreteness, we consider the double-dot set-up shown in Fig.~\ref{fig:setup}. In this set-up a large dot plays the role of the reservoir (R) while the small dot provides the spin (S). We assume that the charging energies of both the large and small dots are large and that both are tuned in the Coulomb blockade regime. In addition, suppose that the level spacing of the small dot is large while that of the large dot, $\Delta$, can be small. Under these conditions, the large dot acts as a finite conduction electron reservoir with a fixed number of electrons, $N$. The leads in Fig.~\ref{fig:setup} are just used to probe the excitation spectrum of the large dot and are assumed to be very weakly coupled.

If the tunnel coupling between the large and small dots is small, we can represent the small dot through a spin ${\mathcal S}=1/2$ operator, and write down a simple model to describe the reservoir-spin system:
\begin{equation}
\label{eq:basic_ham}
H_\text{R-S} = \sum_{\alpha \sigma} \epsilon_\alpha c^{\dagger}_{\alpha \sigma}c_{\alpha\sigma} 
+ {\mathcal J} \;{\mathbf {\mathcal S}}\cdot {\mathbf s} (0)
+ E_C( N-{n_g})^2 \;.
\end{equation}
The first term describes electron-hole excitations in the reservoir (large dot), with the operators $c^{\dagger}_{\alpha\sigma}$ creating one body eigenstates on R with energy $\epsilon_\alpha$, wave function $\phi_\alpha ({\bf r})$ and spin $\sigma$. The second term describes the antiferromagnetic exchange interaction between the two dots, with $ {\bf s}(0) = \frac 1 2 f^{\dagger}_{0\sigma} {\overrightarrow{\sigma}_{\sigma\sigma^\prime}} f_{0\sigma}$ the spin density in the large dot at the tunneling position ${\bf r} \equiv 0$ and $f^{\dagger}_{0\sigma}= \sum_{\alpha} \phi_{\alpha}(0) c^\dagger_{\alpha}$. For completeness we retain with the last term the electrostatic energy of the reservoir; $n_g$ and $E_C$ are the dimensionless gate voltage applied to the large dot and its charging energy.

\textit{Ground state spin of $H_\text{R-S}$---}We now prove an exact theorem for the ground state spin of the system described by Eq.~(\ref{eq:basic_ham}). To simplify the ensuing discussion, we divide $H_{\rm R-S}$ into diagonal and off-diagonal parts
$H_\text{R-S} = H_{\rm D} + H_{\rm OD}$, with
\begin{eqnarray}
\label{eq:dod}
\lefteqn{
 H_{\rm D} = {\mathcal J} {\mathcal S^z} s^z (0) + \sum_i \alpha_i \;f_{i\sigma}^{\dagger } f_{i\sigma}+ E_C( N-{n_g})^2} \\ \nonumber
& & \hspace*{-0.19in}
H_{\rm OD} = \frac{|{\mathcal J}|}{2}\bigl[{\mathcal S^+} s^-(0)+ {\rm h.c.} \bigr]
- \sum_{i,\sigma}\bigl(|t_{i}|\; f_{i\sigma}^{\dagger}f_{i+1\sigma} + {\rm h.c.}\bigr).
 \end{eqnarray}
We first made a unitary transformation to represent the kinetic part of $H_\text{R-S}$ by a one-dimensional chain (tridiagonal matrix). The coefficients $\alpha_i, t_{i}$ and the basis states $f^\dagger_{i\sigma}$ are fixed by the choice of the first basis state, $f^{\dagger}_{0 \sigma}$. Absolute value signs may be used because the signs of the hopping coefficients $t_{i}$ and $\mathcal{J}$ can be fixed arbitrarily by choosing the phases of the $f_{i \sigma}$ and making a $\pi$ rotation about the $z$ axis defined by $\mathcal{S}^z$.

Consider now the following set of many body states that span the Fock space of Eq.~({\ref{eq:basic_ham}}),
\begin{equation}
\label{eq:basis}
| \phi_\alpha \rangle = (-1)^{m-\mathcal{S}}
f_{i_{N_\uparrow}{\uparrow}}^{\dagger}
... 
f_{i_1\uparrow}^{\dagger} 
f_{j_1\downarrow}^{\dagger}
... 
f_{j_{N_\downarrow}{\downarrow}}^{\dagger} 
| 0 \rangle \otimes | m \rangle\;,
\end{equation}
with $m = {\cal S}^z$ and the site labels (positive integers) ordered as $i_1<\dots<i_{N_{\uparrow}}$ and $j_1<\dots<j_{N_{\downarrow}}$.
In this basis $H_{\rm D}$ has only diagonal matrix elements and $H_{\rm OD}$ has only off-diagonal ones. Note that the Hamiltonian conserves both the total number of electrons, $N=N_\uparrow+N_\downarrow$, and the $z$-component of the total magnetization, $S^z_{\rm tot} =(N_\uparrow-N_\downarrow)/2 + m$.

It is easy to show that in this basis all {\em off-diagonal} matrix elements of the Hamiltonian are {\em negative}. Furthermore, repeated application of $H_\text{R-S}$ connects every basis state within a subspace $(S^z_{\rm tot},N)$ of fixed $S^z_{\rm tot}$ and charge $N$. These facts allow us to generalize the proof of the ``Marshall sign theorem'' \cite{Aue} to this case with several implications: the expansion of the ground state in the basis (\ref{eq:basis}) has strictly positive coefficients, and hence the ground state in each $(S^z_{\rm tot},N)$ sector is unique.

These results allow us to make specific statements about the ground state {\em spin} of $H_{\rm R-S}$. First consider ${\cal S}=1/2$. Since the Hamiltonian has SU(2) symmetry, we can label its eigenstates by $S^z_{\rm tot}$ and the total spin $S_{\rm tot}$. For a fixed realization of the reservoir R, we can always make ${\cal J}$ small enough that simple degenerate perturbation is convergent. With $N$ odd, it is then easy to show that the ground state $|G\rangle_{{\cal J} \to 0}$ is a singlet with $S_{\rm tot} = S^z_{\rm tot} =0$. Because both $|G\rangle_{{\cal J} \to 0 }$ and $|G\rangle_{S^z_{\rm tot} = 0, {\cal J}}$, the ground state in sector $S_{\rm tot}^z=0$ at arbitrary ${\cal J}$, can be expanded in the basis (3) with strictly positive coefficients, they must have non-zero overlap, implying that $|G\rangle_{S^z_{\rm tot} = 0, {\cal J}}$ also has total spin $S_{\rm tot}=0$. Furthermore, this state must be the overall ground state, since any multiplet with $S_{\rm tot} > 0 $ necessarily has a state with $S^z_{\rm tot}=0$.

Thus we conclude that the ground state is always a spin singlet for any $N$ odd. These considerations can be easily extended for $N$ even to prove that \textit{the ground state spin of $H_\text{R-S}$ with $\mathcal{S}=1/2$, is always $S_{\rm tot} = 0$ for $N$ odd and $S_{\rm tot} = 1/2$ for $N$ even.}

Note that so far we have made no assumptions regarding the energies $\epsilon_\alpha$ and wave functions $\phi_\alpha(0)$ and so, remarkably, the above theorem holds regardless of how disordered/chaotic the electrons in the reservoir are. We pause here to point out that the theorem applies to arbitrary spin, ${\mathcal S}>1/2$, or ferromagnetic coupling, ${\mathcal J}<0$. It is also straightforward to generalize it to the Anderson model, which then takes into account charge fluctuations on the small dot \cite{TBP}. Interestingly, for ${\mathcal S}=1$ and ${\mathcal J}>0$, we find that the ground state can never be a singlet: for $N$ odd it is a doublet while for $N$ even it is a spin triplet \cite{TBP}. In the ${\mathcal S}=1/2$ ferromagnetic Kondo problem, we find that the ground state is always either ${\mathcal S}=1$ or ${\mathcal S}=1/2$ depending on the parity of $N$.

\textit{Excited states---}In addressing the nature of the excited states, we now specialize to the usual ${\mathcal S}=1/2$, ${\mathcal J}>0$ case, but similar arguments can be constructed for ${\mathcal S}>1/2$ or ${\mathcal J}<0$. For an arbitrary $J/\Delta$ in the ${\mathcal S}=1/2$, ${\mathcal J}>0$ case, Marshall's sign theorem combined with a weak coupling expansion only guarantees that the first excited state is either $S_{\rm tot}=0$ or $1$ ($1/2$ or $3/2$) for the $N$ odd (even) case. Therefore, we shall examine the excited states by perturbation theory in the two limits $\Delta \gg T_{\rm K}$ (${\cal J} \ll \Delta$) and $\Delta \ll T_{\rm K}$ to gain more insight into the structure of the excitation spectrum.

In the weak coupling regime, $\Delta \gg T_{\rm K}$, given a realization of the reservoir R, we can always make ${\cal J}$ small enough that the spectrum can be constructed through lowest order degenerate perturbation theory. For $N$ odd, the ground state is a singlet and the first excited state is a triplet with excitation energy $\delta E = {\mathcal J}|\phi_{\alpha_{\rm top}}(0)|^2 \ll \Delta$, where $\alpha_{\rm top}$ is the topmost singly occupied level of the large dot. The next excited state is a singlet of energy $\sim \Delta$ separated from a triplet by a splitting $\sim \cal J$. For $N$ even, the ground state is a doublet and the first excited state is an 8-fold degenerate multiplet for $\mathcal{J}=0$ that gets split into two $S_{\rm tot} = 1/2$ doublets and one $S_{\rm tot} = 3/2$ quadruplet. In general the two doublets have lower (though unequal) energy than the quadruplet.

For $T_{\rm K} \gg \Delta$, on the other hand, the impurity spin $\cal S$ is screened by the conduction electrons, and we can use Nozi\`eres' ``Fermi-liquid'' theory \cite{Noz}. For $N$ odd (even) we end up effectively with an {\em even (odd) number} of quasiparticles that interact with each other only at the impurity site through a repulsive effective interaction $U_{\rm FL}\sim (\Delta^2/T_{\rm K}) \, n_{\uparrow} \,n_{\downarrow}$. The quasiparticles have the same mean level spacing $\Delta$ as the original electrons, but the spacing between two quasiparticle levels is not simply related to the spacing of the original levels in the chaotic quantum dot. For $N$ odd, we find that the ground state is a singlet (as expected from our theorem), and the excitations start at energy $\sim \Delta$. The first two excitations consist of a spin $S_{\rm tot}=1$ followed by a $S_{\rm tot}=0$ whose energy is split due to the residual quasiparticle interaction by an amount $\sim \Delta^2/T_K$. A similar analysis for the $N$ even case shows that at strong coupling the ground state and first two excitations are all doublets.

\begin{figure}[t]
\includegraphics[width=3.in]{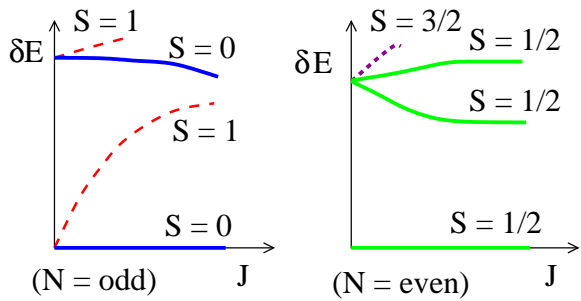}
\caption{(color online) Schematic illustration of the energy eigenvalues of the double dot system Eq.~(\ref{eq:basic_ham}) as a function of the coupling ${\mathcal J}$ for $N$ odd (left) and even (right). In the double-dot experiment proposed in the text, the y-axis is like $V_{\rm BIAS}$ and the x-axis is like $V_{\rm PINCH}$. The trajectory of the excitations will show up as peaks in the differential conductance $G$. 
}
\label{fig:illust}
\end{figure}

Remarkably, the ordering of the $S_{\rm tot}$ quantum numbers of the ground state and two lowest excitations is the same in both the $T_{\rm K} \gg \Delta$ and $T_{\rm K} \ll \Delta$ limits. It is therefore quite natural to assume that the order and quantum numbers are independent of the precise value of $T_K/\Delta$. Thus we arrive at the schematic illustration in Fig.~\ref{fig:illust}.

{\em Scaling for $N$ odd---}To make further progress, we consider the standard simplification: $|\phi_\alpha(0)|^2 =1$ and $\epsilon_{\alpha+1} - \epsilon_{\alpha} = \Delta$ independent of $\alpha$ \cite{ThiSim}. Under these conditions, the spectrum, and thus the splitting between the ground and first excited states, must be a universal function of $T_{\rm K}/\Delta$ that only depends on the {\em parity} of the number electrons on the dot. For $N$ odd, for example, the singlet-triplet gap must scale as $\delta E_{\rm ST}/\Delta = F(T_{\rm K}/\Delta)$. At strong coupling, the behavior $F(x\gg1)\approx1$ follows from the arguments above. To obtain the scaling behavior of $F$ at weak-coupling, we sum the leading logarithmically divergent series in the perturbative expansion of $\delta E_{\rm ST}$ in $\mathcal{J}$; this gives $\delta E_{\rm ST}  = \mathcal{J}/ (1-\frac{\mathcal{J}}{\Delta}\ln(D/\Delta))$, whence $F(x\ll1)\approx1/{\rm ln}(1/x)$.

We find $F(x)$ in the whole crossover regime through continuous time quantum Monte Carlo calculations. The method we use relies on a mapping onto an $XY$ spin chain with an impurity attached to it \cite{Wil,YooCha}. The updates maintain the number of particles (canonical ensemble). Details of the method and additional results shall be presented elsewhere \cite{TBP}.

\begin{figure}[t]
\centering
\includegraphics[width=3in,clip]{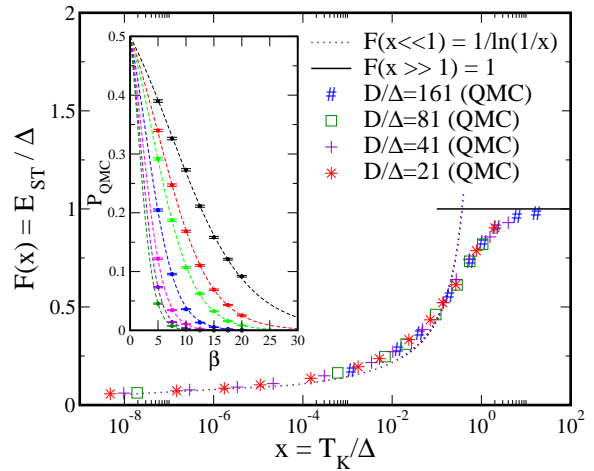}
\caption{(color online) Universal form of the singlet-triplet gap ($N$ odd). Note the excellent data collapse upon plotting the extracted gaps as a function of $T_{\rm K}/\Delta$. Inset: $P_{\rm QMC}(\beta)$ for $D/\Delta=41$. Circles are data for fixed ${\mathcal J}=0.05$, $ 0.10$, $ 0.13$, $ 0.15$, $ 0.19$, $ 0.23$, $ 0.26$, and $ 0.30$ (right to left). Lines are one parameter fits to the two state single-triplet form.}
\label{fig:qmc}
\end{figure}

To extract $\delta E_{\rm ST}$, we measure the fraction $P$ of states with ${S^z_{\rm tot}}^2=1$ visited in the Monte Carlo sampling at temperature $T$. For a fixed ${\mathcal J}$ and large $\beta=1/T$, $P(\beta)$ can be excellently fit to the form ${2}/({ 3 + e^{\beta\, \delta E_{\rm ST}}})$ valid for a two-level singlet-triplet system (inset to Fig.~\ref{fig:qmc}). Repeating this procedure yields $\delta E_{\rm ST}$ for a variety of ${\mathcal J}$ and $\Delta$. Plotted as a function of $T_{\rm K}/\Delta$ (see Fig.~\ref{fig:qmc}), we find a nice data collapse for different combinations of the bare parameters over eight orders of magnitude in $T_{\rm K}/\Delta$. These calculations thus illustrate in this special case that the ground state is a singlet for $N$ odd and that the first excited state remains indeed a triplet for all values of ${\cal J}$.

{\em Exchange coupling---}So far, electron-electron interactions on the large dot have only been incorporated through the last term of Eq.~(\ref{eq:basic_ham}). However, exchange interaction between the electrons also induces a Hund's rule coupling, $H_\text{ex} = -J_{\rm S} \; S_\text{R}^2$ ($S_\text{R}$ is the total electronic spin in the reservoir), which will compete with the Kondo effect \cite{Mur}. The form of this coupling invalidates our exact theorem. However, since the coupling $J_{\rm S}>0$ is typically smaller than the level spacing in semiconductor systems \cite{Oregetal} and its effect (to our knowledge) has not been observed yet in the metallic grains \cite{DelRal}, we expect that it would only slightly modify the predicted spectrum, introducing a small splitting between the states having different spins in Fig.~\ref{fig:illust}. There is a statistical chance that mesoscopic fluctuations of the level spacing will completely destroy this picture for some realizations of R \cite{BarUll}.

\textit{Tunneling spectrum---}In the setup of Fig.~\ref{fig:setup}, for very weak tunneling to the leads, the differential conductance between the leads as a function of bias voltage allows one to extract directly the excitation spectrum of the S-R system in its final state \cite{DelRal}. Assuming, e.g., that the reservoir initially has an even number of electrons, it is possible to measure the $N$ odd excitation spectrum. We have predicted (Fig.~\ref{fig:illust}) that this spectrum changes in an essential way as a function of the pinch-off voltage between the two dots that drives the system from the weakly coupled limit to the strongly correlated Fermi liquid state. A further probe is achieved by applying a parallel magnetic field (Zeeman) that couples to the system via $B_{\rm Z}S_{\rm tot}^z$, thus splitting the spin multiplets.

\begin{figure}[t]
\includegraphics[width=2.6in,clip]{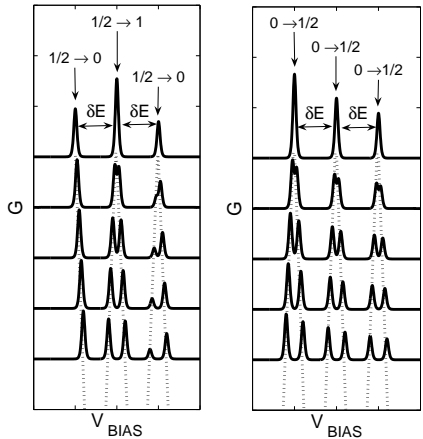}
\caption{Conductance through the S-R system of Fig.~\ref{fig:setup} as a function of the bias between the two leads for ${B_{\rm Z}}=(0,0.05,0.1,0.15,0.2)\delta E$ (from top to bottom). For clarity, all the zero field splittings $\delta E$ of the resonances shown are taken equal. See text for more details. The initial state of the reservoir has $N$ even (odd) for the left (right) panel. Both the total spin and energy splitting (as displayed schematically in Fig.~\ref{fig:illust}) of the device as a function of $\mathcal{J}$ can be extracted from such data.}
\label{fig:diff_cond}
\end{figure}

To find the conductance in such a tunneling experiment, we have solved the rate equations for transferring an electron from lead 1 to the reservoir and then to lead 2 \cite{Beenakker,DelRal}. We assume that (1) the coupling of the lead to each state in R is the same (mesoscopic fluctuations are neglected), (2) the Kondo correlations that develop in S-R do not affect the matrix element for coupling to the leads, (3) there is a transition rate $\lambda_{\rm rel}$ that provides direct thermal relaxation between the eigenstates of S-R with fixed $N$, (4) the electrons in the lead are in thermal equilibrium, and (5) the temperature $T$ is larger than the widths $\Gamma_1,\Gamma_2$ of the S-R eigenstates due to L1 and L2. 
With these assumptions, the rate equations and their solution are standard. 
Results are shown in Fig.~\ref{fig:diff_cond} in the case of parameters: $\Gamma_2/\Gamma_1=0.1$, $T/\delta E=0.1$, and $\lambda_{\rm rel}/\Gamma_2=1$. 
Note in particular two features. First, the first peak in the even$\rightarrow$odd case does not split in field while for odd$\rightarrow$even it does; this provides a definite signature of the parity of $N$. Second, for even$\rightarrow$odd, the behavior of the excited states as a function of field is quite different: the triplet excitation splits into two symmetric peaks while the singlet excitation splits very asymmetrically. Thus the singlet and triplet excitations can be identified unambiguously and so the behavior shown in Fig.~\ref{fig:illust} verified. 

{\em Larger spin---}Our theorem and the analysis of excited states can be easily generalized to the underscreened ${\mathcal S}>1/2$ single channel Kondo problem, which has indeed been realized experimentally \cite{triplet}. Interestingly, for $N$ even and ${\mathcal S}=1$, it follows from our theorem that the ground state for {\em all} $T_{\rm K}/\Delta$ has $S_{\rm tot}=1$. At weak coupling, we deduce this simply from perturbation theory and also that the first excited state is a singlet at energy $\sim\Delta$. In the opposite limit $T_{\rm K}\gg \Delta$, the flow is to the usual underscreened fixed point \cite{Hew}: one is left with a ${\mathcal S}=1/2$ impurity coupled \textit{ferromagnetically} to an odd number of quasiparticles in the reservoir. Since the ferromagnetic Kondo problem flows naturally to weak coupling \cite{Hew}, we are again justified in doing perturbation theory, and we recover that the ground state $S_{\rm tot}=1$ is separated from the first singlet excitation by an asymptotically small energy. Thus, the signature of the underscreened Kondo effect is a \textit{reduction} of the gap between triplet ground and singlet excited states, of logarithmic form as a function of $T_K/\Delta$, as one tunes from weak to strong coupling.

In conclusion, with a combination of an exact theorem, perturbative arguments, and Monte Carlo simulations, we have made specific predictions on the low-lying excitation spectrum for a double dot system and a metallic grain containing a single magnetic impurity. The excitation spectrum was found to reflect the formation of the Kondo-correlated many-body state in an essential way. Further by numerically solving the corresponding rate equations, we have shown that the gate voltage and magnetic field dependencies of conductance measurements can be used to observe our predictions directly in an experiment.

We acknowledge useful discussions with K.A.~Matveev, and A.~Rosch, and thank J.~Yoo for use of his quantum Monte  Carlo code. This research was supported in part by the US NSF grant (DMR-0506953) and Hungarian Grants OTKA Nos.~T046267, T046303 and NF061726.

\end{document}